\documentclass[12pt,prd,showpacs,tightenlines,nofootinbib]{revtex4}
\usepackage{bm}
\usepackage{graphics}
\usepackage{rotating}
\usepackage{epsfig}
\begin{document}
\begin{flushright}{HU-EP-08/68}\end{flushright}
\title{Relativistic model of hidden bottom tetraquarks }
\author{D. Ebert$^{1}$, R. N. Faustov$^{2}$  and V. O. Galkin$^{1,2}$}
\affiliation{
$^1$ Institut f\"ur Physik, Humboldt--Universit\"at zu Berlin,
Newtonstr. 15, D-12489  Berlin, Germany\\
$^2$ Dorodnicyn Computing Centre, Russian Academy of Sciences,
  Vavilov Str. 40, 119991 Moscow, Russia}

\begin{abstract}
The relativistic model of the ground state and excited heavy
tetraquarks with hidden bottom  is formulated within  the
diquark-antidiquark picture.  The diquark structure is taken into
account by calculating the diquark-gluon vertex in terms of the
diquark wave functions. Predictions for the masses of bottom
counterparts to the charm tetraquark candidates are given.   
\end{abstract}


\maketitle

During last few years a significant experimental progress has been achieved in
meson spectroscopy. Many new heavy meson states have been
discovered. Some of them are long-awaited states (such as $h_c$,
$\eta_b$, etc.) while other states (such as $X(3872)$, $Y(4260)$,
$Z(4430)$, etc.) cannot be easily fitted in the simple $q\bar q$ picture of 
mesons \cite{pakhlova}. These anomalous states and especially the charged ones
can be considered as indications of the existence of
exotic multiquark states which were predicted long ago
\cite{jm,bk}. Very recently the Belle Collaboration \cite{belleY} observed an
enhancement in  $e^+e^- \to \Upsilon(1S)\pi^+\pi^-,
\Upsilon(2S)\pi^+\pi^-$, and $\Upsilon(3S)\pi^+\pi^-$ production which
is not well-described by the conventional $\Upsilon(10860)$
line shape. One of the possible explanations is a bottomonium
counterpart to the $Y(4260)$ state which may overlap with the
$\Upsilon(5S)$. New data on higher bottomonium 
excitations are expected to 
come in near future from KEKB, LHC and Tevatron. It is important to
note that it is planned to search for bottom partners of anomalous
charmonium-like states at LHC.

In  papers
\cite{tetr1,tetr2} we calculated masses of the ground and excited
states of heavy tetraquarks in  the framework of the relativistic quark
model based on the quasipotential approach in quantum
chromodynamics. It was found that most of the anomalous
charmonium-like states could be interpreted as the diquark-antidiquark
bound states.
Here we extend this analysis to the consideration of the
excited tetraquark states with hidden bottom.
As previously, we use the  diquark-antidiquark
picture to reduce a complicated relativistic 
four-body problem to the subsequent two more simple two-body
problems. The first step consists in the calculation of the masses, wave
functions and form factors of the diquarks, composed from light and bottom
quarks in the colour antitriplet state. At the second step, a bottom
tetraquark is considered to be a 
bound diquark-antidiquark system. It is 
important to emphasize that the diquark is not  a point
particle but  its structure is explicitly taken into account by calculating
the diquark-gluon vertex.

In the adopted approach the  quark-quark bound state and
diquark-antidiquark bound state  are described
by the diquark wave function ($\Psi_{d}$)  and by the tetraquark wave
function ($\Psi_{T}$), respectively.  These wave functions satisfy the
quasipotential equation of the Schr\"odinger type~\cite{efg}
\begin{equation}
\label{quas}
{\left(\frac{b^2(M)}{2\mu_{R}}-\frac{{\bf
p}^2}{2\mu_{R}}\right)\Psi_{d,T}({\bf p})} =\int\frac{d^3 q}{(2\pi)^3}
 V_{d,T}({\bf p,q};M)\Psi_{d,T}({\bf q}),
\end{equation}
where the relativistic reduced mass is
\begin{equation}
\mu_{R}=\frac{E_1E_2}{E_1+E_2}=\frac{M^4-(m^2_1-m^2_2)^2}{4M^3},
\end{equation}
and $E_1$, $E_2$ are given by
\begin{equation}
\label{ee}
E_1=\frac{M^2-m_2^2+m_1^2}{2M}, \quad E_2=\frac{M^2-m_1^2+m_2^2}{2M}.
\end{equation}
Here, $M=E_1+E_2$ is the bound-state mass (diquark or tetraquark),
$m_{1,2}$ are the masses of quarks ($q$ and $Q$) which form
the diquark or of the diquark ($d$) and antidiquark ($\bar d'$) which
form the heavy tetraquark ($T$), and ${\bf p}$ is their relative
momentum. In the center-of-mass system the relative momentum
squared on mass shell reads
\begin{equation}
{b^2(M) }
=\frac{[M^2-(m_1+m_2)^2][M^2-(m_1-m_2)^2]}{4M^2}.
\end{equation}
The kernel $V_{d,T}({\bf p,q};M)$ in Eq.~(\ref{quas}) is the
quasipotential operator of the quark-quark or diquark-antidiquark
interaction. It is constructed with the help of the off-mass-shell
scattering amplitude, projected onto the positive-energy states.
The explicit expressions for the corresponding quasipotentials
$V_{d,T}({\bf p,q};M)$  can be 
found in Ref.~\cite{tetr2}.

The constituent quark masses $m_b=4.88$ GeV,
$m_u=m_d=0.33$ GeV, $m_s=0.5$ GeV and the parameters of the linear
potential $A=0.18$ GeV$^2$ and $B=-0.3$~GeV have been fixed previously
and have values typical in quark models. The value of the mixing coefficient of
vector and scalar confining potentials $\varepsilon=-1$ has been
determined from the consideration of charmonium radiative decays
\cite{efg} and the heavy-quark expansion. The universal
Pauli interaction constant $\kappa=-1$ has been fixed from the
analysis of the fine splitting of heavy quarkonia ${ }^3P_J$ -
states \cite{efg}. In this case, the long-range chromomagnetic
interaction of quarks vanishes in accordance with the flux-tube
model.

At the first step, we take the previously calculated  masses and form
factors of the bottom diquarks \cite{tetr1}.  The
diquark interaction with the gluon field, which 
takes into account the diquark structure, is expressed through the
form factor $F(r)$ entering the vertex of the 
diquark-gluon interaction \cite{hbar}. This form factor is determined
through the overlap integral of the diquark wave functions. Our
estimates showed that this form factor can be approximated  with a 
high accuracy by the expression 
\begin{equation}
  \label{eq:fr}
  F(r)=1-e^{-\xi r -\zeta r^2}.
\end{equation}
The values of the masses and parameters $\xi$ and $\zeta$ for the bottom
scalar diquark $[\cdots]$ and axial vector diquark $\{\cdots\}$ ground
states are given in Table~\ref{tab:dmass}.

\begin{table}
  \caption{Masses $M$ and form factor  parameters of bottom
    diquarks \cite{tetr1}. $S$ and $A$ 
    denote scalar and axial vector diquarks which are antisymmetric $[\cdots]$ and
    symmetric $\{\cdots\}$ in flavour, respectively. }
  \label{tab:dmass}
\begin{ruledtabular}
\begin{tabular}{ccccc}
Quark& Diquark&   $M$ &$\xi$ & $\zeta$
 \\
content &type & (MeV)& (GeV)& (GeV$^2$)  \\
\hline
$[b,q]$& $S$ & 5359 &6.10 & 0.55 \\
$\{b,q\}$& $A$ & 5381& 6.05 &0.35 \\
$[b,s]$ & $S$& 5462 & 5.70 &0.35 \\
$\{b,s\}$& $A$ & 5482 & 5.65 &0.27
  \end{tabular}
\end{ruledtabular}
\end{table}

At the second step, we calculate the masses of heavy tetraquarks 
considered as the bound states of the bottom diquark and
antidiquark. 
The explicit expression for the 
diquark-antidiquark interaction is given in \cite{tetr2}.
In this picture of heavy tetraquarks
both scalar $S$ (antisymmetric in flavour
$[Qq]_{S=0}=[Qq]$) and axial vector $A$ (symmetric in flavour
$[Qq]_{S=1}=\{Qq\}$) diquarks are considered.
As a result, a very rich set of tetraquark states
emerges. However the number of states in the considered
diquark-antidiquark picture is significantly less than in the genuine
four-quark approach.

\begin{table}
  \caption{Masses of hidden bottom tetraquark  ground ($1S$) states
    (in MeV) \cite{tetr1}. $S$ and $A$ 
    denote scalar and axial vector diquarks. }
  \label{tab:bmass}
\begin{ruledtabular}
\begin{tabular}{ccccc}
State& Diquark &
\multicolumn{3}{l}{\underline{\hspace{3.5cm}Mass\hspace{3.5cm}}} 
\hspace{-5.5cm} \\
$J^{PC}$ & content& $bq\bar b\bar q$ &$bq\bar b\bar s$ & $bs\bar b\bar
s$ \\
\hline
$1S$\\
$0^{++}$ & $S\bar S$ & 10471 & 10572 & 10662\\
$1^{+\pm}$ & $(S\bar A\pm \bar S A)/\sqrt2$& 10492 & 10593& 10682\\
$0^{++}$& $A\bar A$ & 10473 & 10584& 10671\\
$1^{+-}$& $A\bar A$ & 10494 & 10599& 10686\\
$2^{++}$& $A\bar A$ & 10534 & 10628& 10716\\
 \end{tabular}
\end{ruledtabular}
\end{table}

\begin{table}
  \caption{Thresholds for open bottom decays.}
  \label{tab:bthr}
\begin{ruledtabular}
\begin{tabular}{cccccc}
Channel& Threshold (MeV)&Channel& Threshold (MeV)&Channel& Threshold
(MeV)\\ 
\hline
$B\bar B$ & 10558 & $B B_s$ &10649& $B_s^+B_s^-$ &10739\\
$B\bar B^*$ & 10604  &$B^* B_s$ & 10695&$B_s^\pm B_s^{*\mp}$ & 10786\\
$B^*\bar B^*$ & 10650  &$B^* B_s^*$& 10742 &$B_s^{*+} B_s^{*-}$ & 10833 
\end{tabular}
\end{ruledtabular}
\end{table}

The previously calculated  
masses of the tetraquark ground ($1S$) states \cite{tetr1} 
and the corresponding open bottom  thresholds are shown in
Tables~\ref{tab:bmass}, \ref{tab:bthr}. Note that most of the
tetraquark states were predicted (in contrast to tetraqurks with hidden charm
\cite{tetr1}) to lie significantly
 below corresponding open bottom
thresholds. The mass of the bottom counterpart to $X(3872)$ is predicted
to be 10492~MeV. 
In Tables~\ref{tab:ecmass},~\ref{tab:ecmass1} we give predictions for the
orbitally and radially excited tetraquark states with hidden
bottom. Excitations only in the diquark-antidiquark system are
considered.

\begin{table}
  \caption{Masses of hidden bottom tetraquark excited  $1P$, $2S$ states
     (in MeV).  $S$ and $A$
    denote scalar and axial vector diquarks; $\cal S$ is the total
    spin of the diquark and antidiquark. ($C$ is defined only for
    neutral states).}
  \label{tab:ecmass}
\begin{ruledtabular}
\begin{tabular}{cccccc}
State& Diquark & &
\multicolumn{3}{l}{\underline{\hspace{2.9cm}Mass\hspace{2.9cm}}} 
\hspace{-5.5cm} \\
$J^{PC}$ & content&$\cal S$& $bq\bar b\bar q$ &$bq\bar b\bar s$ & $bs\bar b\bar s$ \\
\hline
$1P$\\
$1^{--}$ & $S\bar S$& 0 & 10807& 10907& 11002\\
$0^{-\pm}$ & $(S\bar A\pm \bar S A)/\sqrt2$&1&10820 & 10917&11011 \\
$1^{-\pm}$ & $(S\bar A\pm \bar S A)/\sqrt2$&1&10824 &10922& 11016 \\
$2^{-\pm}$ & $(S\bar A\pm \bar S A)/\sqrt2$&1&10834 &10932 &11026 \\
$1^{--}$& $A\bar A$ &0&10850& 10947& 11039\\
$0^{-+}$& $A\bar A$ & 1&10836 &10934& 11026\\
$1^{-+}$& $A\bar A$ & 1& 10847 &10945& 11037\\
$2^{-+}$& $A\bar A$ & 1& 10854 & 10952& 11044\\
$1^{--}$& $A\bar A$ & 2&10827 &10925& 11017\\
$2^{--}$& $A\bar A$ & 2& 10856& 10953& 11046\\
$3^{--}$& $A\bar A$ & 2& 10858& 10956& 11048\\
 $2S$\\
$0^{++}$ & $S\bar S$ & 0 & 10917 &11018&11111\\
$1^{+\pm}$ & $(S\bar A\pm \bar S A)/\sqrt2$& 1& 10939 & 11037& 11130\\
$0^{++}$& $A\bar A$ & 0 & 10942 &11041& 11133\\
$1^{+-}$& $A\bar A$ & 1 & 10951 &11050&11142\\
$2^{++}$& $A\bar A$ & 2 & 10969& 11067& 11159\\
 \end{tabular}
\end{ruledtabular}
\end{table}

\begin{table}
  \caption{Masses of hidden bottom tetraquark excited  $1D$, $2P$ states
    (in MeV). $S$ and $A$
    denote scalar and axial vector diquarks; $\cal S$ is the total
    spin of the diquark and antidiquark. }
  \label{tab:ecmass1}
\begin{ruledtabular}
\begin{tabular}{cccccc}
State& Diquark & &
\multicolumn{3}{l}{\underline{\hspace{2.9cm}Mass\hspace{2.9cm}}} 
\hspace{-5.5cm} \\
$J^{PC}$ & content&$\cal S$& $bq\bar b\bar q$ &$bq\bar b\bar s$ & $ bs\bar b\bar s$ \\
\hline
$1D$\\
$2^{++}$ & $S\bar S$& 0 & 11021& 11121& 11216\\
$1^{+\pm}$ & $(S\bar A\pm \bar S A)/\sqrt2$&1&11040 &11137& 11232\\
$2^{+\pm}$ & $(S\bar A\pm \bar S A)/\sqrt2$&1&11042 &11139& 11235\\
$3^{+\pm}$ & $(S\bar A\pm \bar S A)/\sqrt2$&1&11045 &11142& 11238\\
$2^{++}$& $A\bar A$ &0&11064& 11162&11255\\
$1^{+-}$& $A\bar A$ & 1&11060 &11158&11251\\
$2^{+-}$& $A\bar A$ & 1& 11064&11161&11254\\
$3^{+-}$& $A\bar A$ & 1& 11066&11164& 11257\\
$0^{++}$& $A\bar A$ & 2&11054 &11152& 11245\\
$1^{++}$& $A\bar A$ & 2&11057& 11155& 11248\\
$2^{++}$& $A\bar A$ & 2& 11062 & 11159& 11252\\
$3^{++}$& $A\bar A$ & 2& 11066 &11164& 11257\\
$4^{++}$& $A\bar A$ & 2&11067& 11165& 11259\\
$2P$\\
$1^{--}$ & $S\bar S$& 0 & 11122& 11221& 11316\\
$0^{-\pm}$ & $(S\bar A\pm \bar S A)/\sqrt2$&1&11134 & 11232&11326\\
$1^{-\pm}$ & $(S\bar A\pm \bar S A)/\sqrt2$&1&11139 & 11236&11330\\
$2^{-\pm}$ & $(S\bar A\pm \bar S A)/\sqrt2$&1&11148 & 11245&11340\\
$1^{--}$& $A\bar A$ &0&11163& 11260&11353\\
$0^{-+}$& $A\bar A$ & 1&11151& 11248& 11342\\
$1^{-+}$& $A\bar A$ & 1&11161 &11259& 11351\\
$2^{-+}$& $A\bar A$ & 1&11168& 11265& 11358\\
$1^{--}$& $A\bar A$ & 2&11143& 11241& 11333\\
$2^{--}$& $A\bar A$ & 2&11169 &11266& 11359\\
$3^{--}$& $A\bar A$ & 2&11172 &11269& 11362\\
 \end{tabular}
\end{ruledtabular}
\end{table}

Our model predicts three
vector $1^{--}$ tetraquark states with hidden bottom in the mass range
10807--10850~MeV. One of these tetraquarks is composed from a scalar
diquark and antidiquark ($S\bar S$). Two other $1^{--}$ states contain
an axial vector diquark and antidiquark with the total spin of the
diquark and antidiquark $\cal S$ equal to 0 and 2. Two
lighter tetraquarks, with predicted masses 10807~MeV ($S\bar S$) and
10827~MeV ($A\bar A$), are bottom partners of $Y(4260)$ while the
heavier one, with mass 10850~MeV ($A\bar A$), is the bottom partner of
$Y(4360)$. Therefore a complicated structure of vector bottomonium
states emerges in this mass range, which can be responsible for 
the anomalous production cross sections for $e^+e^-\to
\Upsilon(1S,2S,3S)\pi^+\pi^-$ observed by Belle \cite{belleY}. 
Their fit using a single Breit-Wigner resonance shape yielded a peak
mass of $10889.6\pm 1.8\pm 1.5$~MeV. However,  a more
detailed experimental study is necessary to clarify this question. The
bottom counterpart to the vector state $Y(4660)$ has predicted mass
11122~MeV.  It is very important to search for the charged or strange states with
hidden bottom. Their observation will be a
direct proof of the existence
of heavy tetraquarks. The masses of the bottom counterparts to charged
$Z(4248)$ 
and  $Z(4430)$ are predicted at around 10807~MeV and
10939~MeV, respectively. 

It is 
necessary to emphasize that the observation of the bottom
counterparts to the new anomalous charmonium-like states is very
important since it will allow to discriminate between different
theoretical descriptions of these states. Indeed theoretical models,
such as the hybrid, molecular or diquark-antidiquark pictures etc., give 
significantly different results in the bottom sector (see also
discussions in \cite{hou,kl}).

In summary, we obtained predictions for the masses of tetraquarks with 
hidden bottom in the diquark-antidiquark picture. For the calculations
we used the dynamical approach based on the relativistic quark model
which was previously successfully applied in the charm sector. The
tetraquark  
masses were obtained by numerical solution of the quasipotential wave
equation with the corresponding relativistic 
potentials. The diquark structure was taken into account in terms of
diquark wave functions. In our analysis we did not introduce any free
adjustable 
parameters but used their values fixed from the previous considerations
of heavy and light meson properties \cite{efg}. The correspondence
between bottom  and charm tetraquark candidates 
was discussed. 

The authors are grateful to  V. Matveev,
G. Pakhlova and  V. Savrin  for support and discussions.
 One of us
(V.O.G.) thanks M. M\"uller-Preussker and the colleagues 
from the particle theory group for kind hospitality. 
This work was supported in part by 
Deutscher Akademischer Austauschdienst (DAAD) (V.O.G.),
the Russian Science Support Foundation 
(V.O.G.) and the Russian Foundation for Basic Research (RFBR), grant
No.08-02-00582 (R.N.F.~and~V.O.G.).

\end{document}